\documentclass[aps,pra,showpacs,twocolumn,floatfix,amsmath,preprintnumbers,nofootinbib,letter,secnumarabic]{revtex4} 

\usepackage[utf8]{inputenc}
\usepackage{color}
\usepackage{graphicx,epsfig,verbatim}
\usepackage{amssymb,amsfonts}
\usepackage{hyperref}

\usepackage[caption=false]{subfig}

\def\art{paper}
\def\jrn#1#2#3#4#5#6{#3 \textbf{#4}, #5 (#6)} \def\andd{, and } 

\def\scn#1#2{\section{#1}\lb{#2}}
\def\eq{eq.\,} \def\eqs{eqs.\,} 

\def\bfl{\begin{flushleft}}
\def\efl{\end{flushleft}}
\def\bfr{\begin{flushright}}
\def\efr{\end{flushright}}
\def\bc{\begin{center}}
\def\ec{\end{center}}
\def\be{\begin{equation}}
\def\ee{\end{equation}}
\def\bse{\begin{subequations}}
\def\ese{\end{subequations}}
\def\ba{\begin{eqnarray}}
\def\ea{\end{eqnarray}}
\def\baa#1{\begin{array}{#1}}
\def\eaa{\end{array}}
\def\bw{\begin{widetext}}
\def\ew{\end{widetext}}
\def\nn{\nonumber }
\def\lb#1{\label{#1}}
\def\bit{\begin{itemize}}
\def\eit{\end{itemize}}
\def\bco{}
\def\bcs{\begin{cases}}
\def\ecs{\end{cases}}

\def\Der#1#2{\frac{\drm #1}{\drm #2}}

\def\vol{V}

\def\nrmf{{N}}

\def\Mass{{\cal M}}
\def\Mass{{M}}
\def\lan{{\cal L}}
\def\lanp{{\cal V}}
\def\en{{E}}
\def\vena{\boldsymbol{\nabla}}

\def\drm{d}
\def\dvol{\drm\vol}

\def\dn{\rho}  \def\dnc{\bar{\dn}}
\def\en{\epsilon}
\def\vel{\vartheta}

\begin{document}

\preprint{\small  Low Temp. Phys. \textbf{45}, 1231 (2019) 
\quad 
[\href{https://doi.org/10.1063/10.0000200}{DOI: 10.1063/10.0000200}]
}

\title{
Resolving the puzzle of 
sound propagation in liquid helium 
at low temperatures
}

\author{Tony C. Scott}
\affiliation{Institut f\"{u}r Physikalische Chemie, RWTH Aachen University, Aachen 52056, Germany}
\affiliation{BlockFint, 
139 Sethiwan Tower, 4A Pan Rd, 
Bangkok 10500, Thailand}

\author{Konstantin G. Zloshchastiev}
\email{kostiantynz@dut.ac.za}
\affiliation{Institute of Systems Science, Durban University of Technology, P.O. Box 1334, Durban 4000, South Africa}

\begin{abstract}
Experimental data suggests that, at temperatures below 1 K, the pressure in liquid helium
has a cubic dependence on density. Thus the speed of 
sound scales as a cubic root of pressure.
Near a critical pressure point, this speed approaches zero
whereby the critical pressure is negative, thus indicating a cavitation instability regime.
We demonstrate that to explain this dependence, one has 
to view liquid helium as a mixture of three 
quantum Bose liquids:
dilute (Gross-Pitaevskii-type) Bose-Einstein condensate, 
Ginzburg-Sobyanin-type fluid,
and logarithmic superfluid.
Therefore, the dynamics of such a mixture is described by a quantum wave equation, 
which contains not only the 
polynomial (Gross-Pitaevskii and Ginzburg-Sobyanin) nonlinearities
with respect to a condensate wavefunction, but also a non-polynomial logarithmic nonlinearity.
We derive an equation of state and speed of sound in our model, and show their agreement with experiment.
\end{abstract}

\date{received: 
6 Mar 2019 [APS], 7 Jun / 20 Oct 2019 [LTP]; published: 31 Dec 2019 [LTP]
}

\pacs{03.75.Hh, 67.25.dt\\
Keywords: superfluid helium, quantum Bose liquid, equation of state, speed of sound
}

\maketitle

\begin{figure}[h]
\epsfig{figure=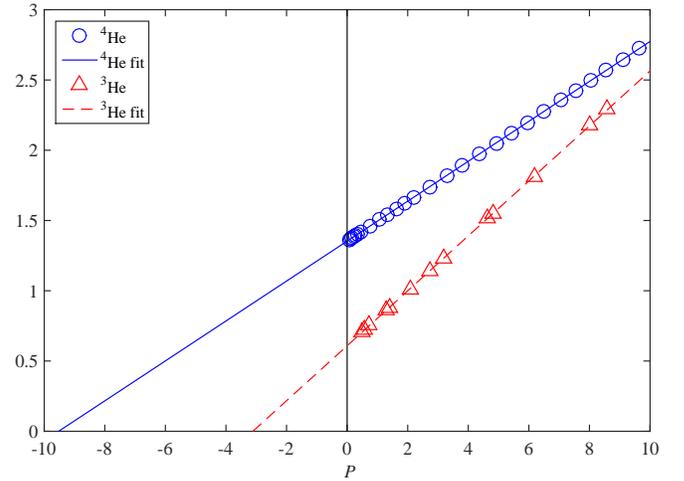,width=  0.99\columnwidth}
\caption{
Profile of $c_s^3$ in units of $10^{13} \text{cm}^3/\text{s}^3$, versus pressure $P$, in atm.
The experimental data was taken from Refs. \cite{abraham1,abraham2}, for $^4$He (circles, solid fitting curve) and $^3$He 
(triangles, dashed fitting curve). 
}
\label{f:fexpsnd}
\end{figure}

\scn{Introduction}{s:intro}
The velocity of ordinary (``first'') sound in liquid helium at temperatures below 1 K 
was measured with great accuracy as a function of both temperature \cite{wc62,wc67}
and pressure \cite{abraham1,abraham2}.
An analysis of experimental data by Abraham \textit{et al.}
reveals that sound velocity $c_s$
decreases with pressure
as a cubic root, cf. Fig. \ref{f:fexpsnd}:
\be\lb{e:correct}
c_s = K^{1/3} (P - P_c )^\nu,
\ 
\nu =
\tfrac{1}{3} \pm 0.01
, 
\ee
where the critical pressure 
being
about $P_c = -9.52$ and $-3.11$ atm for $^4 \text{He}$ and $^3 \text{He}$ respectively,
whereas $ K = 1.41 \times 10^6$ $\text{cm}^4\, \text{g}^{-1} \,\text{s}^{-1}$
and $1.93 \times 10^6$ $\text{cm}^4\, \text{g}^{-1} \,\text{s}^{-1}$ for $^4 \text{He}$
and $^3 \text{He}$ respectively;
negative values indicate that zero velocity of sound
occurs in the cavitation
regime where nucleation of bubbles causes a macroscopic instability.
On the other hand, conventional arguments imply that 
in the vicinity of the critical pressure point $P_c$
the speed of sound should scale as a quartic root of pressure,
$
c \propto (P - P_c )^{1/4} 
$,
which disagrees with both experiment 
and numerical simulations \cite{mar91,mar94,mar95,Qu,bauer,dalfovo,boronat,edwards}.
These arguments rely heavily on approximations and perturbation techniques,
which are usually pertinent to systems of weakly interacting bosons,
such as dilute Bose-Einstein condensates, where two-body contact interactions are predominant
and non-perturbative effects are neglected.

In this \art, we propose a 
non-perturbative approach, which takes into account multi-body interactions in liquid helium
and explains the experimental data
by Abraham \textit{et al.}, including an equation of state and the relation \eqref{e:correct}.

\scn{Model}{s:mod}
%
After the discovery of Bose-Einstein condensation and related phenomena, it
was established that quantum Bose liquids are not a discrete set of particles, such
as helium atoms, interacting
via some interparticle potential. 
Instead, a new phenomenon occurs:
the discreteness of the
atoms is averaged out,
collective degrees of freedoms emerge, which are no longer
identical to 
constituent particles,
and the whole system becomes essentially nonlocal and continuous \cite{ant03}.
Correspondingly, wavefunctions
describing these new degrees of freedom are to be considered fundamental
within the frameworks of the collective approach. 
Even for ground states,
these condensate wavefunctions do not obey the conventional (linear) Schr\"odinger equations,
but some nonlinear quantum
wave equations in which nonlinearities account for many-body effects \cite{gs76,gs82,jpr95}. 
Our task here will be to
determine a form of those equations, starting from some minimal model
assumptions followed by fixing their parameters using experimental data. 

In a theory of quantum Bose liquids and Bose-Einstein condensates,
one introduces a complex-valued condensate wavefunction
$\Psi = \Psi (\textbf{x},t)$,
which 
obeys a normalization condition
\be\lb{e:norm}
\int_\vol |\Psi|^2 \dvol  
= \int_\vol \dn \dvol 
= \Mass = m \nrmf 
,\ee 
where $\Mass$, $\vol$ and $\nrmf$
are, respectively, the total mass, volume, and particles' number of the fluid,
$\dn = |\Psi|^2$ is the fluid mass density,
and $m =  m_{\text{He}}$ is the mass of the constituent particle.
This condition imposes restrictions 
upon the condensate wavefunctions,
which reveals its quantum mechanical nature:
the set of all normalizable functions $\Psi$ 
which constitutes a Hilbert space, such as $L^2$.

For physical configurations,
the  function $\Psi$ must minimize the 
action functional
$\iint_\vol \lan \, \dvol \drm t$,
where the
Lagrangian density has a Galilean-invariant $U(1)$-symmetric form:
\be\lb{e:ftlan}
\lan
= 
\frac{i \hbar}{2}(\Psi \partial_t\Psi^* - \Psi^*\partial_t\Psi)+
\frac{\hbar^2}{2 m}
|\vena \Psi|^2
+ V_\text{ext}\, |\Psi|^2
+
\lanp (|\Psi|^2)
,
\ee
where $V_\text{ext} = V_\text{ext} (\textbf{x},t)$ is an external or trapping potential,
and $\lanp (|\Psi|^2)$ is an effective many-body interaction potential density.
We write the latter in the series form, to be explained as follows:
\ba
\lanp (\dn)
&=& 
\lanp_{(\text{ln})} (\dn) 
+
\sum\limits_{k=2}^{\cal N}
\lanp_{(k)} (\dn) 
,\lb{e:ftpot}\\
\lanp_{(\text{ln})}  (\dn) 
&=& 
\en
C_{(\text{ln})}
\dn 
\left[
1-
\ln{\left(
\dn/\dnc
\right)}
\right]
,
\lb{e:ftpotln}\nn\\
\lanp_{(k)}  (\dn) 
&=& 
\en
C_{(k)}
\dnc
\left(
\dn/\dnc
\right)^{k}
,
\lb{e:ftpotk}\nn
\ea
where $\en$ and $\dnc$ are scale constants with the dimensions of energy and mass density, respectively,
and the
$C$'s are dimensionless coupling coefficients whose values are to be established below;
the index $k$ labels the $k$th-order contact interaction potential with respect to $\dn$.
The couplings $C$'s do not depend on the mass density $\dn$, but they can be functions 
of the thermodynamic parameters of the fluid.

By applying the Euler-Lagrange variational principle to \eqs \eqref{e:ftlan}, \eqref{e:ftpot},
we obtain the following quantum wave equation:
\be\lb{e:becgeneq}
\left[
- i \hbar \, \partial_t
- \frac{\hbar^2}{2 m} \vec \nabla^2
+
V_{\text{ext}} (\vec x,\, t)
+
F(|\Psi|^2)
\right]
\Psi
= 0,
\ee
where 
\ba
F(\dn) &=&
\Der{}{\dn}
\lanp (\dn)
=
F_{(\text{ln})} (\dn) 
+
\sum\limits_{k=2}^{\cal N}
F_{(k)} (\dn) 
,
\lb{e:F} \\ 
F_{(\text{ln})}  (\dn) 
&=& 
-
\en
C_{(\text{ln})}
\ln{\!\left(
\dn/\dnc
\right)}
,
\lb{e:Fln}\nn\\
F_{(k)}  (\dn) 
&=& 
\en
k
C_{(k)}
\left(
\dn/\dnc
\right)^{k-1}
.
\lb{e:Fk}\nn
\ea
Equations \eqref{e:ftpot}-\eqref{e:F} indicate that the Bose liquid in our model has the following multi-component structure:

First, the potential term 
$\lanp_{(\text{ln})} \propto
|\Psi|^2
\left[
1-
\ln{\left(
|\Psi|^2/\dnc
\right)}
\right]
$
describes the logarithmic fluid component and induces the logarithmic nonlinearity 
$F_{(\text{ln})}  \propto
\ln{\left(
|\Psi|^2/\dnc
\right)}
$ in the wave equation \eqref{e:becgeneq}.
Nonlinearity of this type often occurs in field theories of particles and gravity \cite{ros68,ros69,bb76,em98,z10gc,z11appb,dz11,scott1,dmz15,scott2},
as well as
in the classical and quantum mechanics of various fluids \cite{z11appb,dmf03,gl08,az11,z12eb,bo15,z17zna,z18epl,z18zna}. 
The reason for such universality is that logarithmically nonlinear terms readily emerge in evolution equations for those dynamical systems in which interparticle interaction energies dominate kinetic ones \cite{z18zna}.
Such systems include not only low-temperature gases and liquids \cite{az11,z12eb}, but also hot dense matter and effectively lower-dimensional flows \cite{dmf03,z18epl}.

In particular, 
a significant amount of experimental evidence supports universal applicability
of logarithmic models in the theory of superfluidity of $^4$He.
The logarithmic fluid turns out to be very instrumental for describing the
microscopic properties of the 
superfluid component of $^4$He:
it analytically reproduces with high accuracy the three main 
observable facts of this liquid -- the Landau spectrum of excitations,
the structure factor, and the speed of sound at normal pressure,
while using only one non-scale parameter to fit
the excitation spectrum's experimental data \cite{z12eb}.

It should be mentioned here that the logarithmic nonlinearity readily occurs
if one attempts to renormalize perturbative models of liquid helium
and take into account zero-point oscillations therein \cite{mar94}.
From that prospective, the logarithmic term can be interpreted as a cumulative macroscopical effect
of the quantum interaction between the collective degrees of freedom of superfluid helium
and virtual particles,
which correlates with an idea of using the logarithmic fluids for a non-perturbative description of
the physical vacuum as such \cite{z11appb}.

Second, the component described by the Ginzburg-Landau-type (quartic) potential $\lanp_{(2)}  \propto |\Psi|^4$,
represents a well-known Gross-Pitaevskii (GP) condensate where the
interparticle interaction can be well approximated by a 2-body potential made of a contact (delta-singular) shape \cite{gr61,pi61}.
This approximation is robust in dilute Bose-Einstein condensates, but for strongly-interacting quantum liquids, it cannot be used as a leading-order approximation.  However, it can still make a viable contribution. 

Third, 
the fluid described by the sextic term, $\lanp_{(3)}  \propto |\Psi|^6$,
is another beyond-GP-approximation term,
as discussed by Ginzburg and Sobyanin \cite{gs76,gs82}.
This term is related to three-particle interactions:
for instance, it can be induced by the trimer bound states in liquid helium
discovered by Efimov \cite{ef71,ef73}, a recent review of 
literature can be found in \cite{kms11}.
Another source of six-order terms with respect to $|\Psi|$
could be fermionic admixtures \cite{cr04,crt04} and 
the reduction of a fluid's effective dimensionality \cite{ks92}.

Finally, the higher-order polynomial potential terms $\lanp_{(k)}  \propto |\Psi|^{2 k}$, where $k > 3$,
can also occur in strongly-interacting Bose liquids.
However, their substantial influence on the physics of liquid helium has, to the best of our knowledge, not been reported yet. 
Below it will be demonstrated that such terms can be neglected for the purposes of this \art.

Using the Madelung representation of a wavefunction,
one can rewrite  any nonlinear wave equation of the type \eqref{e:becgeneq} in hydrodynamic form.
One can show that the corresponding fluid 
has an equation of state and speed of sound $c_s$,
which are given by the following general formulae  
\cite{z11appb}:
\ba
P (\dn) &\approx& P_0
-\frac{\hbar}{m}
\int \dn 
\,\drm F (\dn)
,\lb{e:eosgen}\\
c_s
&=&
\sqrt{\Der{P(\dn)}{\dn}}
\approx
\sqrt{
-\frac{\hbar\, \dn}{m}
\Der{F (\dn)}{\dn}
}
,
\lb{e:csgen}
\ea
where $P_0$ is an arbitrary constant,
and
the approximation symbol means
that we keep only the leading-order terms with respect to the Planck constant;
a detailed derivation of these formulae can be found, \textit{e.g.}, in Sec. 3.1 of \cite{z19mat}.

When evaluated on a function \eqref{e:F}, 
formulae \eqref{e:eosgen} and \eqref{e:csgen} yield
an equation of state and speed of sound for our model:
\ba
P 
&\approx& P_0
+\frac{\hbar \en}{m}
\left[
C_{(\text{ln})} \dn
-
\dnc
\sum\limits_{k=2}^{\cal N}
(k-1)
C_{(k)}
\left(
\dn/\dnc
\right)^{k}
\right]\!,~~~ \lb{e:eos0}\\
c_s^2
&\approx&
\frac{\hbar \en}{m}
\left[
C_{(\text{ln})} 
-
\dnc
\sum\limits_{k=2}^{\cal N}
k
(k-1)
C_{(k)}
\left(
\dn/\dnc
\right)^{k-1}
\right]\!.
\lb{e:cs0}
\ea
Notice here that the logarithmic nonlinearity induces a linear term in the equation of state
and a constant term in an expression for a speed of sound squared,
which confirms the earlier results \cite{z11appb}. 
Therefore,  if one regards \eqref{e:eos0} as a series expansion of a general function
$P (\dn) $ then the logarithmic fluid component
provides a first-order approximation, which corresponds to an ideal fluid.

In the next section,
our aim will be to further specify and narrow the model \eqref{e:ftlan}, \eqref{e:ftpot},
by fixing values of its parameters $C_{(\text{ln})}$ and $C_{(k)}$
to fit the experimental data.
If some of those parameters turn out to be zero then the corresponding 
component in our model can be safely neglected.

\begin{figure}[t]
\centering
\subfloat[$\alpha < 0$]{
  \includegraphics[width=0.49\columnwidth]{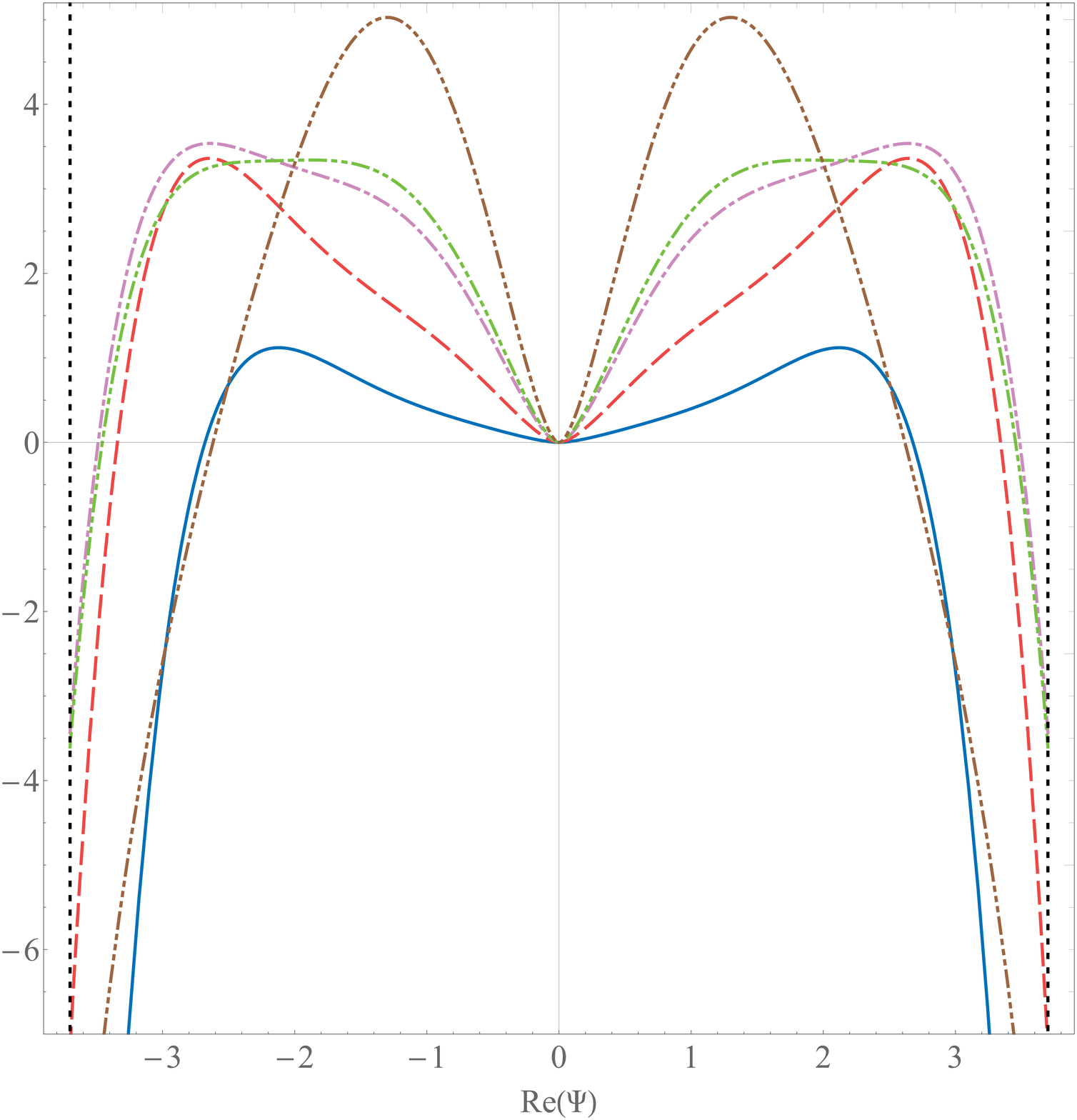}
}
\subfloat[$\alpha > 0$]{
  \includegraphics[width=0.49\columnwidth]{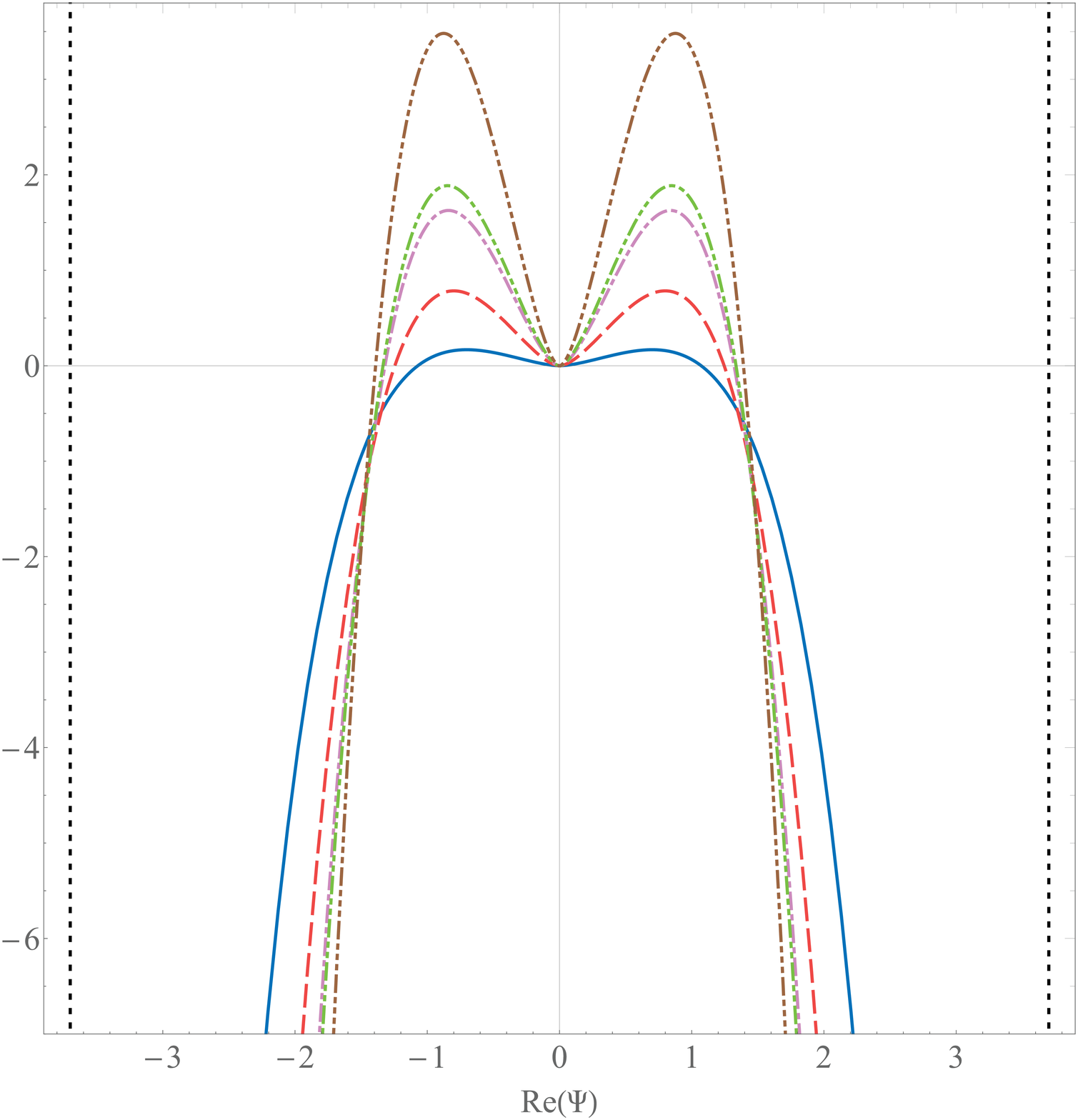}
}
\hspace{0mm}
\caption{Potential 
\eqref{e:ftpotA} divided by $\en\dnc$, versus 
$\text{Re}(\Psi)$ in units of  $\sqrt{\dnc}$, for the following values
of $\alpha$: $\pm 0.5$ (solid curve), $\pm 1$ (dashed),  $\pm 1.4$ (dash-dotted), $\pm 1.5$ (dash-double-dotted),
$\pm 2$ (dash-triple-dotted).
Vertical dotted lines represent an infinite well occurring due to the condition 
$|\Psi| \leqslant |\Psi_\text{max}| < \infty$, which follows from  
the wavefunction's normalization \eqref{e:norm}.
}
\label{f:fpot}
\end{figure}

\scn{Theory vs experiment}{s-exp}
%
In order to fit the experimental data of Ref. \cite{abraham1},
we compare their empirical formulae for an
equation of state and a speed of sound with our equations \eqref{e:eos0} and \eqref{e:cs0}.
Since their empirical equation of state is a cubic polynomial,
one can immediately deduce that our model 
must be truncated at a $k=3$ term:
\be
{\cal N} = 3 \ \Leftrightarrow \ C_{(k)} = 0, \  
\forall
\, k>3
, 
\ee
and we derive constraints for the remaining coefficients.
One can show that those coefficients can be rewritten in the form
\be
C_{(\text{ln})} = \alpha^2
, \
C_{(2)} = - \alpha/3
, \
C_{(3)} = - 1/(2 \ 3^3)
,
\ee
where $\alpha$ is a real constant, which
indicates that the couplings $C_{(\text{ln})}$ and $C_{(2)}$ are not independent
from each other, as experimental data suggest.
This particular choice of coefficients also ensures that the speed
of sound and the pressure difference $P-P_c$ share a single common
real root in $\dnc$, as we would anticipate from \eq \eqref{e:correct}.

Correspondingly, \eqs \eqref{e:ftpot} and \eqref{e:F} become:
\ba
\lanp (\dn)
&=& 
\lanp_{(\text{ln})} (\dn) 
+
\lanp_{(2)} (\dn) 
+
\lanp_{(3)} (\dn) 
,\lb{e:ftpotA}\\
\lanp_{(\text{ln})}  (\dn) 
&=& 
\en
\alpha^2
\dn 
\left[
1-
\ln{\left(
\dn/\dnc
\right)}
\right]
,
\nn\\
\lanp_{(2)}  (\dn) 
&=& 
-
\frac{1}{3}
\en\alpha\dnc
\left(
\dn/\dnc
\right)^{2}
,
\nn\\
\lanp_{(3)}  (\dn) 
&=& 
-
\frac{1}{2\, 3^3}
\en\dnc
\left(
\dn/\dnc
\right)^{3}
,
\nn
\ea
and 
\be\lb{e:Fexp}
F(\dn) =
-
\en
\left[
\alpha^2 \ln{(
\dn/\dnc
)} 
+ \frac{2 \alpha}{3}  \frac{\dn}{\dnc} 
+ \frac{1}{18} \frac{\dn^2}{\dnc^2}
\right]
. 
\ee
One can immediately verify that the total many-body potential density \eqref{e:ftpotA} is regular in
a finite domain, and has a typical Mexican-hat shape which opens either up or down, depending
on the sign of $\en$, cf. Fig. \ref{f:fpot}.
This indicates the possibility of spontaneous symmetry breaking, which indeed occurs in the theory
of both Gross-Pitaevskii condensates and logarithmic fluids, as discussed in Refs. \cite{z11appb,az11,z12eb,z18epl}.

Note that the many-body potentials $\lanp_{(2)} $ and $\lanp_{(3)} $
can be either repulsive or attractive here, depending on the sign of $\en\alpha$;
but the potential $\lanp_{(\text{ln})}$ can switch between attractive and repulsive behaviors
when the fluid density crosses the value $e \dnc$, where $e$ being the base of the natural logarithm.
This property is essentially the one giving logarithmic fluid the majority of its important features mentioned above.

\begin{figure}[t]
\centering
\subfloat[$|\alpha | = 1/100$]{
  \includegraphics[width=0.49\columnwidth]{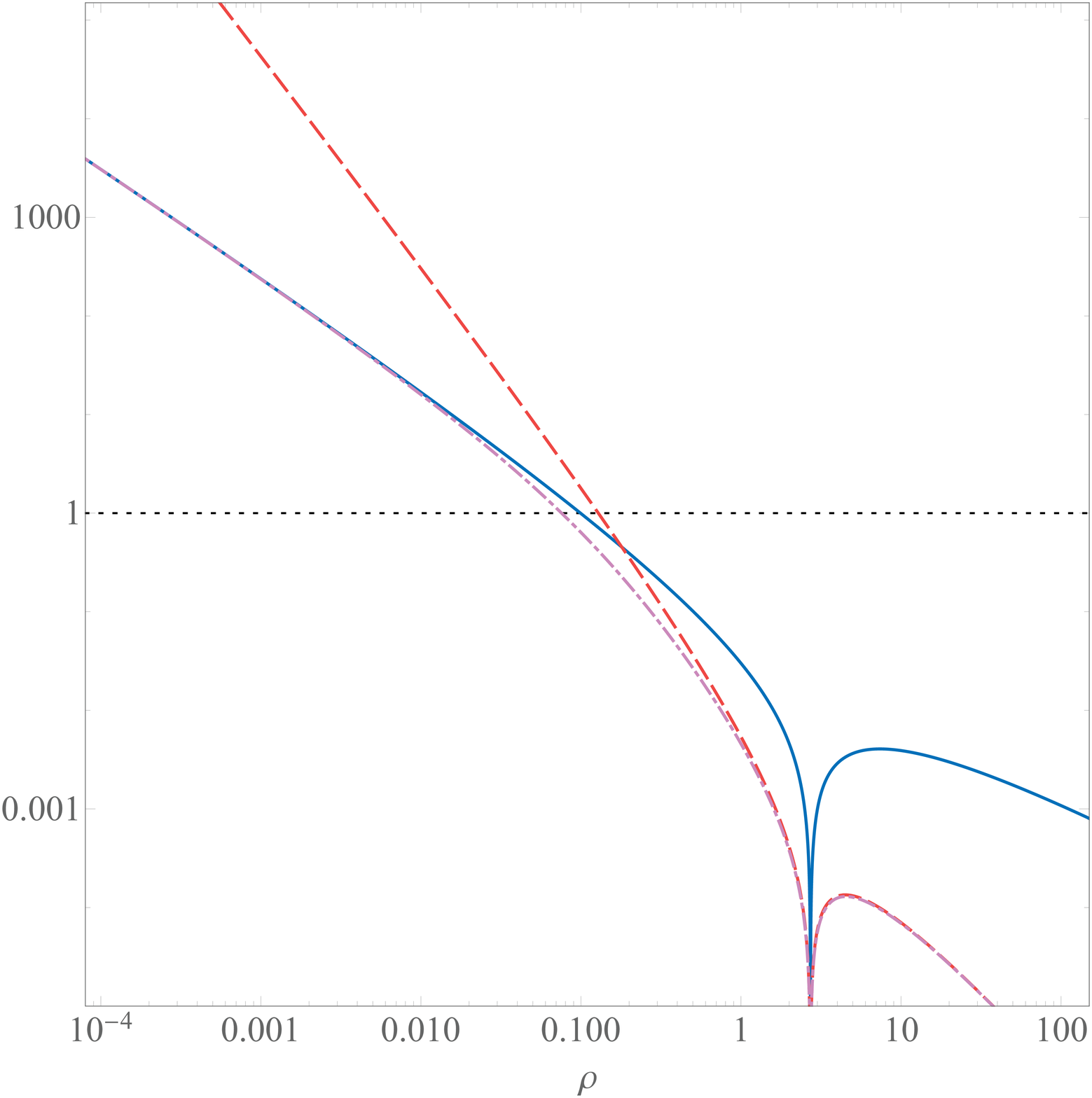}
}
\subfloat[$|\alpha | = 1/2$]{
  \includegraphics[width=0.49\columnwidth]{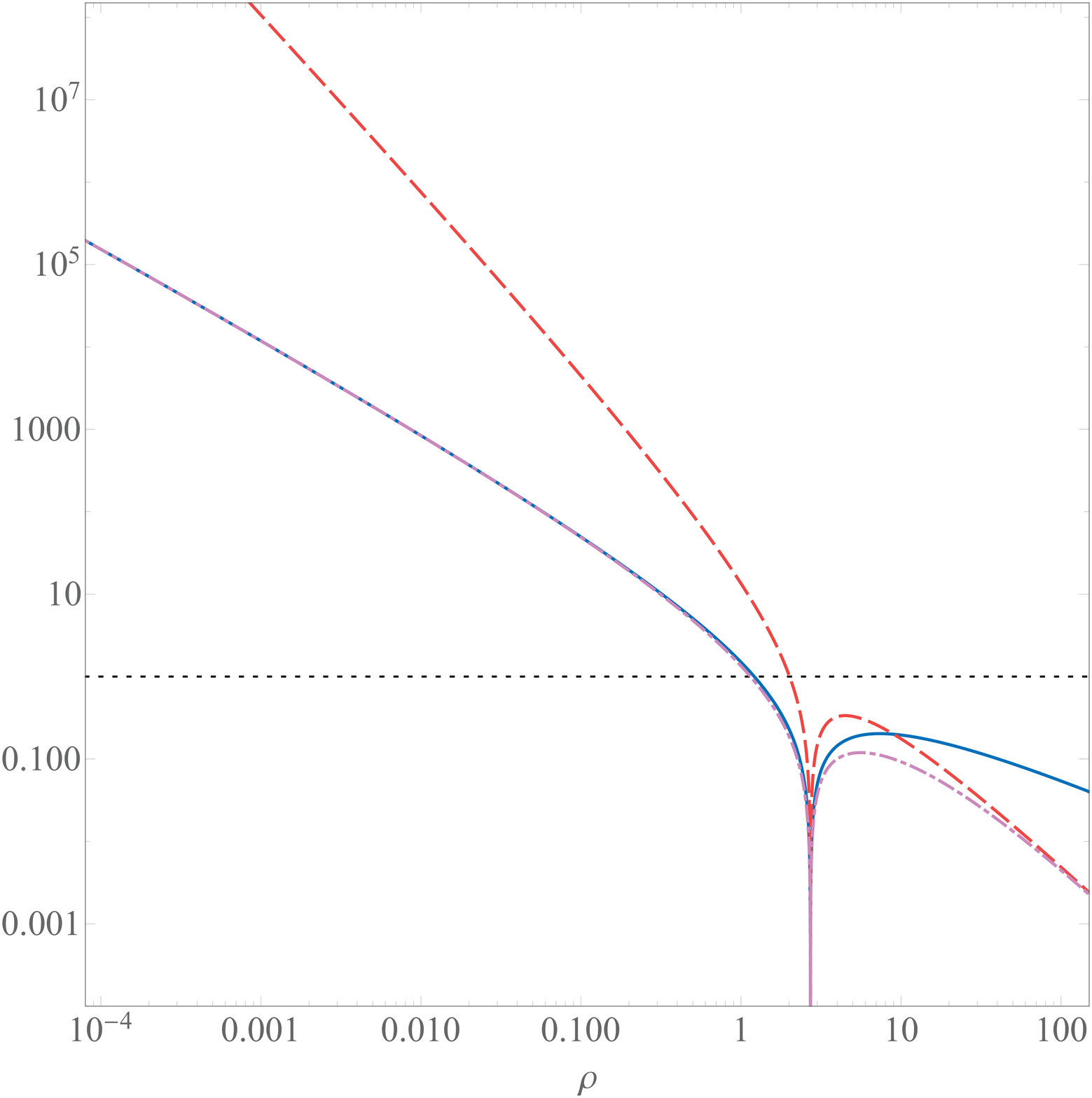}
}
\hspace{0mm}
\subfloat[$|\alpha | = 1$]{
  \includegraphics[width=0.49\columnwidth]{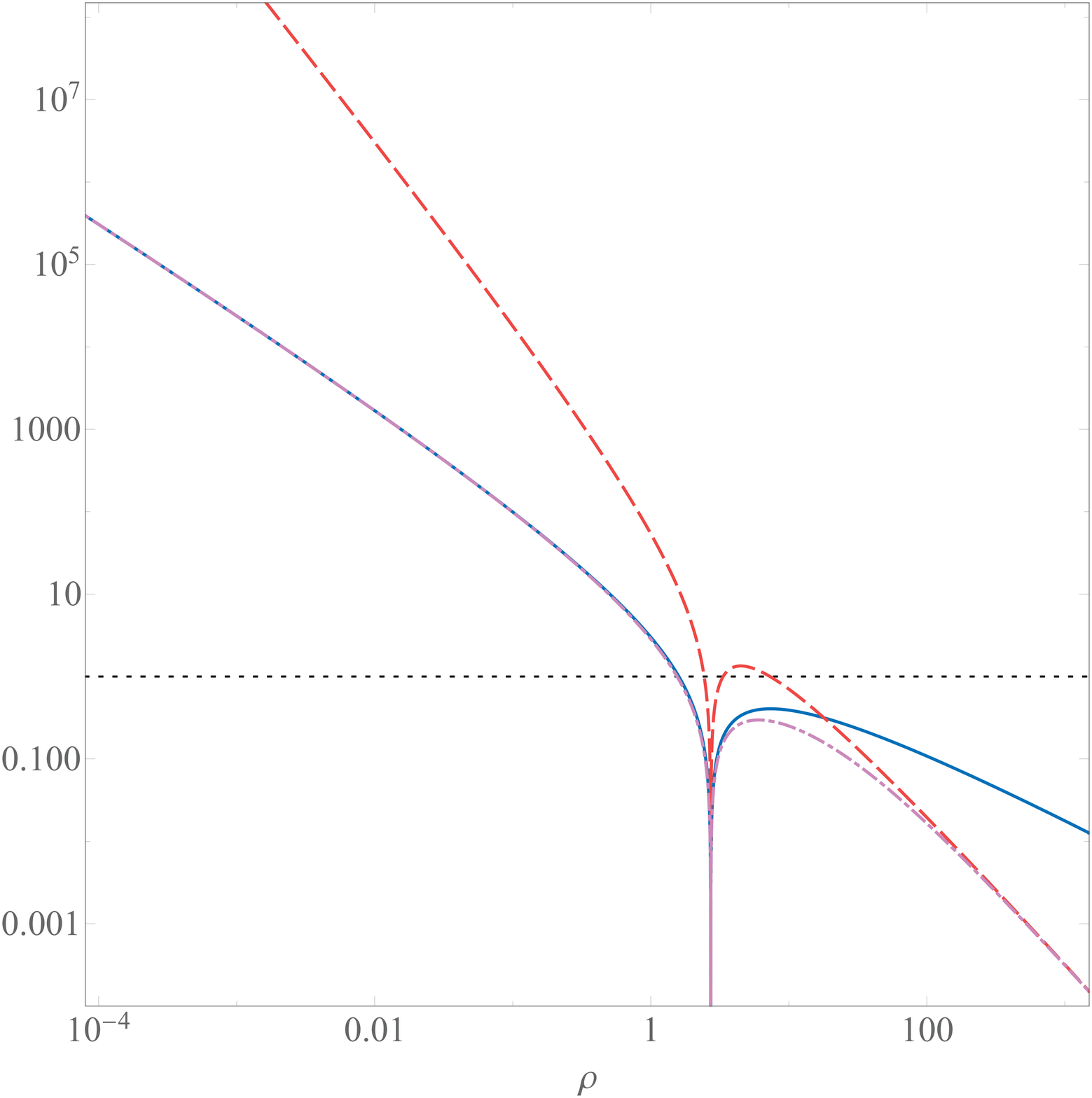}
}
\subfloat[$|\alpha | = 100$]{
  \includegraphics[width=0.49\columnwidth]{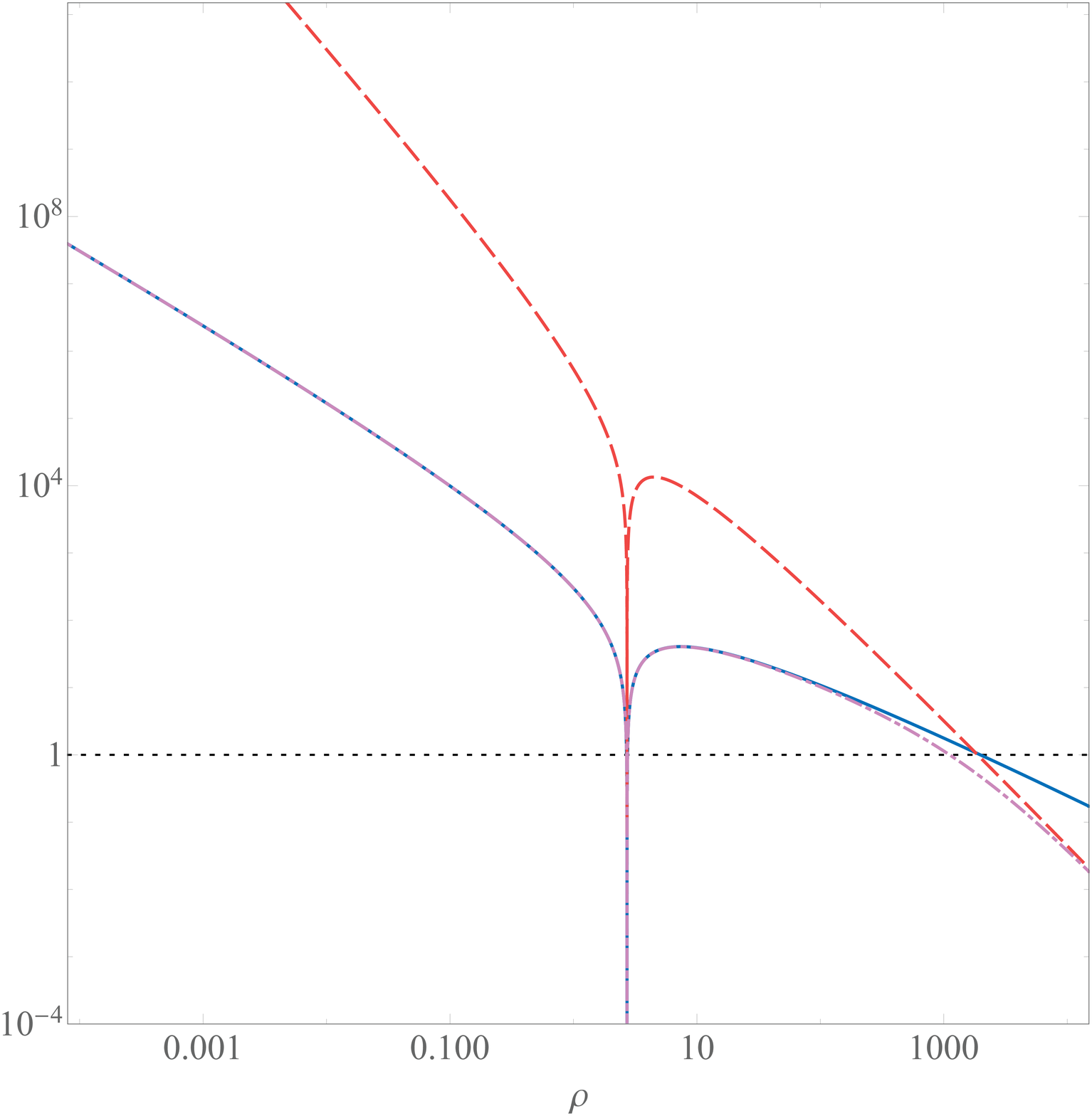}
}
\hspace{0mm}
\caption{Relative values of different terms in a potential \eqref{e:ftpotA}, versus 
density in units $\dnc$: 
$|\lanp_{(\text{ln})}|/|\lanp_{(2)}|$ (solid curve), 
$|\lanp_{(\text{ln})}|/|\lanp_{(3)}|$ (dashed),  
and 
$|\lanp_{(\text{ln})}|/|\lanp_{(2)}+\lanp_{(3)}|$ (dash-dotted).
The horizontal dotted line indicates an equality.
}
\label{f:frat}
\end{figure}

Figure \ref{f:frat} illustrates how much components $\lanp_{(\text{ln})}$,
$\lanp_{(2)} $ and $\lanp_{(3)} $ contribute at different values of density.
One can see that the logarithmic component predominates over the polynomial
ones for quite a large range of density values, but there is always a threshold density, above which  
polynomial terms catch up with and overtake the logarithmic term.
However, this threshold only exists if its value does not exceed a maximum density value occurring due to the condition 
$\dn \leqslant |\Psi_\text{max}|^2 < \infty$;
the latter follows from  
\eq \eqref{e:norm}.
The domination of the logarithmic component explains,
along with its natural applicability for condensate-like systems \cite{z18zna}, 
why the logarithmic fluid is so robust as 
a leading-order approximation model for a superfluid component of helium II \cite{z12eb}.

Using \eqs \eqref{e:eosgen} -
\eqref{e:Fexp}, one can 
easily obtain the equation of state 
and speed of sound:
\be\lb{e:eos}
P \approx P_c 
+
\dnc \, \vel^2
\left(
\alpha
+
\frac{1}{3} \frac{\dn}{\dnc}
\right)^3\!, 
\ 
c_s 
\approx
\vel
\left(
\alpha
+
\frac{1}{3}
\frac{\dn}{\dnc}
\right)
,
\ee
where $\vel = \sqrt{\en/
m 
}$ is a characteristic velocity scale.

Note that in the vicinity of its critical pressure point, liquid helium
can undergo a cavitation-type phase transition, which is similar 
to the above-mentioned topological structure transition in logarithmic superfluids.
Therefore, the total many-body potential \eqref{e:ftpotA} flips,
which corresponds to a value $\en$ turning negative.
In that case, the speed of sound
can formally take imaginary values, which indicates the occurrence of an instability regime. 


Furthermore,
by eliminating the density $\dn$ from formulae \eqref{e:eos},
we recover the experimentally measured behaviour
of the speed of sound $c_s$ in \eqref{e:correct} with $\nu ={\textstyle \frac{1}{3}}$
where
$ 
K = 
\vel/\dnc
$.
Moreover, in the case of $^4$He,
formulae \eqref{e:eos} 
allow us to assign values to
the following combinations
of hitherto free parameters:
\be
\frac{\vel}{\dnc} 
=
\frac{1}{\dnc}
\sqrt{
\frac{\en}{m}
}
=
K
=
3 \sqrt 3 \sqrt{A_3}
,\ 
\alpha\dnc
=
\frac{1}{3}
\left(
\frac{A_1}{A_2} - \rho_0
\right)
,
\ee
where the values of $A$'s and $\dn_0$ for ${}^4$He are given
in Refs. \cite{abraham1,mar94}:
$A_1 = 5.6\times 10^2$ atm cm$^3$ g$^{-1}$, 
$A_2 =1.097 \times 10^4$ atm cm$^6$ g$^{-2}$, 
$A_3 = 7.33 \times 10^4$ atm cm$^9$ g$^{-3}$, and $\dn_0 = 0.14513$ g cm$^{-3}$,
while $\alpha$ remains an unfixed dimensionless parameter of the model.
This means that our model still has some flexibility left,
because $\alpha$ can be used to fit our model for other experiments if necessary.
In this paper, $\alpha$ can be set to any negative number, which reflects a scale nature of the parameter $\dnc$.

We therefore obtain
$\vel = -4.27\, \alpha^{-1} \times 10^{4} $ $\text{cm} \, \text{s}^{-1}$
and
$\dnc = -\frac{1}{3 \alpha} \rho_c \approx -3.01\, \alpha^{-1} \times 10^{-2} \, \text{g} \,  \text{cm}^{-3}$,
where $\rho_c$ is the 
density corresponding to the critical negative pressure $P_c$;
thus, the ratio
$\vel/\dnc = K \approx 1.41 \times 10^6$ 
$\text{cm}^4\, \text{g}^{-1} \,\text{s}^{-1}$
does not depend on $\alpha$.  
These numerical values finalize the fitting of the model \eqref{e:ftlan}, \eqref{e:ftpotA} for the case of $^4$He;
the case of bosonized liquid $^3$He can be done by analogy.

\scn{Conclusion}{s:con}
A fundamental sound velocity pressure dependence holding up for superfluid helium at low temperatures
was examined in the context of the nonlinear wave equation approach for a fluid's wavefunction.
We demonstrated that to explain this dependence, one has 
to view liquid helium as a mixture of three 
quantum Bose liquids:
a dilute (Gross-Pitaevskii-type) Bose-Einstein condensate, 
a Ginzburg-Sobyanin-type fluid,
and a logarithmic superfluid,
where the latter two components can occur due to 
topological and non-perturbative quantum effects,
Efimov trimer states, and fermionic admixtures.

Consequently, the dynamics of such systems is described by a quantum wave equation, 
which contains both the 
polynomial (cubic and quintic) nonlinearities
with respect to a condensate wavefunction, and an essential non-polynomial logarithmic nonlinearity.
Using the hydrodynamic representation of Schr\"odinger-type equations
and experimental data by Abraham \textit{et al.},
we constrained the hitherto free parameters of this multi-component fluid model,
and theoretically reproduced the
empirical formulae for an equation of state and speed of sound.\\

%


\begin{acknowledgments}
T.C.S. would like to thank 
M. 
Therani
of EngKraft LLC, 
S. 
Kanakakorn of BlockFint Thailand, and 
A. 
L\"{u}chow of the Institut f\"{u}r Physikalische Chemie at RWTH-Aachen University, 
for their support. K.G.Z.'s research is supported by Department of Higher Education and Training
of South Africa
and in part by National Research Foundation of South Africa.
Proofreading of the manuscript by P. Stannard is greatly appreciated.
\end{acknowledgments}


\begin{thebibliography}{99}

\bibitem{wc62}
W. M. Whitney and C. E. Chase,
\jrn{Whitney W M and Chase C E}{Velocity of Sound in Liquid Helium at Low Temperatures}{Phys. Rev. Lett.}{9}{243}{1962}

\bibitem{wc67}
W. M. Whitney and C. E. Chase,
\jrn{Whitney W M and Chase C E}{Ultrasonic Velocity and Dispersion in Liquid Helium II
from 0.15 to 1.8 K}{Phys. Rev.}{158}{200}{1967}

\bibitem{abraham1}
B. M. Abraham, Y. Eckstein, J. B. Ketterson, M. Kuchnir\andd P. R. Roach,
\jrn{Abraham B M, Eckstein Y, Ketterson J B, Kuchnir M\andd Roach P R}{Velocity of Sound, Density, and Gruneisen Constant in Liquid 4He}{Phys. Rev. A}{1}{250-257}{1970}

\bibitem{abraham2}
B. M. Abraham, D. Chung, Y. Eckstein, J. B. Ketterson\andd P. R. Roach,
\jrn{Abraham B M, Chung D, Eckstein Y, Ketterson J B\andd Roach P R}{Sound propagation, density, and viscosity in liquid 3He}{J. Low Temp. Phys.}{6}{521-528}{1972}

\bibitem{mar91}
H. Maris,
\jrn{Maris H}{Critical Phenomena in ${}^3$He and ${}^4$He at $T=0$ K}{Phys. Rev. Lett.}{66}{45}{1991}

\bibitem{mar94}
H. Maris,
\jrn{Maris H}{Nucleation of Bubbles on Quantized Vortices in Helium-4}{J. Low Temp. Phys.}{94}{125}{1994}


\bibitem{mar95}
H. Maris,
\jrn{Maris H}{Theory of Quantum Nucleation of Bubbles in Liquid Helium}{J. Low Temp. Phys.}{98}{403-424}{1995}



\bibitem{Qu}
A. Qu, A. Trimeche, J. Dupont-Roc, J. Grucker\andd Ph. Jacquier,
\jrn{Qu A, Trimeche A, Dupont-Roc J, Grucker J\andd Jacquier Ph}{Cavitation Density of Superfluid Helium-4 around 1 K}{Phys. Rev. B}{91}{214115}{2015}

\bibitem{bauer}
G. H. Bauer, D. M. Ceperley\andd N. Goldenfeld,
\jrn{Bauer G H, Ceperley D M\andd Goldenfeld N}{Path-integral monte carlo simulation of helium at negative pressures}{Phys. Rev. B}{61}{9055–9060}{2000}

\bibitem{dalfovo}
F. Dalfovo, A. Lastri, L. Pricaupenko, S. Stringari\andd J. Treiner,
\jrn{Dalfovo F, Lastri A, Pricaupenko L, Stringari S\andd Treiner J}{Structural and dynamical properties of superfluid helium: A density-functional approach}{Phys. Rev. B}{52}{1193–1209}{1995}

\bibitem{boronat}
J. Boronat, J. Casulleras\andd J. Navarro,
\jrn{Boronat J, Casulleras J\andd Navarro J}{Monte Carlo calculations for liquid He-4 at negative pressure}{Phys. Rev. B}{50}{3427–3430}{1994}

\bibitem{edwards}
H. Maris and D. O. Edwards,
\jrn{Maris H\andd Edwards D O}{Thermodynamic properties of superfluid He-4 at negative pressure}{J. Low Temp. Phys.}{129}{1}{2002}


\bibitem{ant03}
I. N. Adamenko, K. E. Nemchenko and I. V. Tanatarov,
\jrn{Adamenko I N, Nemchenko K E\andd Tanatarov I V}{Application of the theory of continuous media to description of thermal excitations}{Phys. Rev. B}{67}{104513}{2003}


\bibitem{gs76}
V. L. Ginzburg and A. A. Sobyanin, 
\jrn{Ginzburg V L\andd Sobyanin A A}{Superfluidity of helium II near the lambda-point}{Sov. Phys. Usp.}{19}{773}{1976}


\bibitem{gs82}
V. L. Ginzburg and A. A. Sobyanin, 
\jrn{Ginzburg V L\andd Sobyanin A A}{Cavitation versus Vortex Nucleation in a Superfluid Model}{J. Low Temp. Phys.}{49}{507}{1982}


\bibitem{jpr95}
C. Josserand, Y. Pomeau\andd S. Rica,
\jrn{Josserand C, Pomeau Y\andd Rica S}{Cavitation versus Vortex Nucleation in a Superfluid Model}{Phys. Rev. Lett.}{75}{3150}{1995}




\bibitem{ros68}
G. Rosen,
\jrn{Rosen G}{Particlelike solutions to nonlinear complex scalar field theories with positive‐definite energy densities}{J. Math. Phys.}{9}{996}{1968}

\bibitem{ros69}
G. Rosen,
\jrn{Rosen G}{Dilatation Covariance and Exact Solutions in Local Relativistic Field Theories}{Phys. Rev.}{183}{1186}{1969}


\bibitem{bb76}
I.~Bialynicki-Birula and J.~Mycielski,
\jrn{Bialynicki-Birula I\andd Mycielski J}{Nonlinear wave mechanics}{Ann. Phys. (N. Y.)}{100}{62}{1976}


\bibitem{em98}
K. Enqvist and J. McDonald,
\jrn{Enqvist K\andd McDonald J}{Q-Balls and Baryogenesis in the MSSM}{Phys. Lett. B}{425}{309-321}{1998}


\bibitem{z10gc}
K.~G.~Zloshchastiev, 
\jrn{Zloshchastiev K G}{Logarithmic nonlinearity in theories of quantum gravity:
Origin of time and observational consequences}{Grav. Cosmol.}{16}{288}{2010}

\bibitem{z11appb}
K.~G.~Zloshchastiev,
\jrn{Zloshchastiev K G}{Spontaneous symmetry breaking and mass generation as built-in phenomena in logarithmic nonlinear quantum theory}{Acta Phys. Polon.}{42}{261}{2011}
	
	
\bibitem{dz11}
V.~Dzhunushaliev and K.~G.~Zloshchastiev,
\jrn{Dzhunushaliev V\andd Zloshchastiev K G}{Singularity-free model of electric charge in physical vacuum: Non-zero spatial extent and mass generation}{Central Eur. J. Phys.}{11}{325}{2013}

\bibitem{scott1}
T. C. Scott, X. Zhang, R. B. Mann\andd G. J. Fee,
\jrn{Scott T C, Zhang X, Mann R B\andd Fee G J}{Canonical reduction for dilatonic gravity in 3+1 dimensions}{Phys. Rev. D}{93}{084017}{2016}

\bibitem{dmz15}
V.~Dzhunushaliev, A. Makhmudov\andd K.~G. Zloshchastiev,
\jrn{Dzhunushaliev V, Makhmudov A\andd Zloshchastiev K G}{Singularity-free model of electrically charged fermionic particles and gauged Q-balls}{Phys. Rev. D}{94}{096012}{2016}

\bibitem{scott2} 
T. C. Scott and J. Shertzer,
\jrn{Scott T C\andd Shertzer J}{Solution of the logarithmic Schr\"{o}dinger equation with a Coulomb potential}{J. Phys. Commun.}{2}{075014 }{2018}


	
	
	
\bibitem{az11}
A.~V.~Avdeenkov and K.~G.~Zloshchastiev,
\jrn{Avdeenkov A V\andd Zloshchastiev K G}{Quantum Bose liquids with logarithmic nonlinearity: Self-sustainability and emergence of spatial extent}{J.\ Phys.\ B: At. Mol. Opt. Phys.}{44}{195303}{2011}

\bibitem{z12eb}
K.~G.~Zloshchastiev, 
\jrn{Zloshchastiev K G}{Volume element structure and roton-maxon-phonon excitations in superfluid helium beyond the Gross-Pitaevskii approximation}{Eur. Phys. J. B}{85}{273}{2012}


\bibitem{bo15}
B. Bouharia,
\jrn{Bouharia B}{Stability of logarithmic Bose-Einstein condensate in harmonic trap}{Mod. Phys. Lett. B}{29}{1450260}{2015}

\bibitem{z17zna}
K.~G.~Zloshchastiev, 
\jrn{Zloshchastiev K G}{Stability and metastability of trapless Bose-Einstein condensates and quantum liquids}{Z. Naturforsch. A}{72}{677}{2017}


\bibitem{z18zna}
K.~G.~Zloshchastiev, 
\jrn{Zloshchastiev K G}{On the dynamical nature of nonlinear coupling of logarithmic quantum wave equation, Everett-Hirschman entropy and temperature}{Z. Naturforsch. A}{73}{619}{2018}



\bibitem{dmf03}
S. De Martino, M. Falanga, C. Godano\andd G. Lauro,
\jrn{De Martino S, Falanga M, Godano C\andd Lauro G}{Logarithmic Schr\"odinger-like equation in magma}{Europhys. Lett.}{63}{472}{2003}

\bibitem{gl08}
G. Lauro,
\jrn{Lauro G}{A note on a Korteweg fluid and the hydrodynamic form of the logarithmic Schr\"odinger equation}{Geophys. Astrophys. Fluid Dyn.}{102}{373}{2008}

\bibitem{z18epl}
K.~G.~Zloshchastiev, 
\jrn{Zloshchastiev K G}{Nonlinear wave-mechanical effects in Korteweg fluid magma transport}{Europhys. Lett. (EPL)}{122}{39001}{2018}



\bibitem{gr61} 
E. P. Gross,
\jrn{Gross E P}{Structure of a quantized vortex in boson systems}{Nuov. Cim.}{20}{454–457}{1961}


\bibitem{pi61} 
L. P. Pitaevskii,
\jrn{Pitaevskii L P}{Vortex Lines in an Imperfect Bose Gas}{Sov. Phys. JETP}{13}{451}{1961}






\bibitem{ef71}
V. Efimov, 
\jrn{Efimov V}{Weakly-bound states of three resonantly-interacting particles}{Sov. J. Nucl. Phys.}{12}{589}{1971}

\bibitem{ef73}
V. Efimov, 
\jrn{Efimov V}{Energy levels of three resonantly interacting particles}{Nucl. Phys. A}{210}{157}{1973}




\bibitem{kms11} 
E. A. Kolganova, A. K. Motovilov\andd W. Sandhas,
\jrn{Kolganova E A, Motovilov A K\andd Sandhas W}{The$~^4$He trimer as an Efimov system}{Few-Body Syst.}{51}{249-257}{2011}




\bibitem{cr04}
S. T. Chui and V. N. Ryzhov, 
\jrn{Chui S T\andd Ryzhov V N}{Collapse transition in mixtures of bosons and fermions}{Phys. Rev. A}{69}{043607}{2004}

\bibitem{crt04}
S. T. Chui, V. N. Ryzhov\andd E. E. Tareyeva, 
\jrn{Chui S T, Ryzhov V N\andd Tareyeva E E}{Stability of Bose system in Bose-Fermi mixture with attraction between bosons and fermions}{JETP Lett.}{80}{274–279}{2004}



\bibitem{ks92}
E. B. Kolomeisky and J. P. Straley, 
\jrn{Kolomeisky E B\andd Straley J P}{Low-Dimensional Bose Liquids: Beyond the Gross-Pitaevskii Approximation}{Phys. Rev. B}{46}{11749}{1992}


\bibitem{z19mat}
K.~G.~Zloshchastiev, 
\jrn{Zloshchastiev K G}{Matrix logarithmic wave equation and multi-channel systems in fluid mechanics}{J. Theor. Appl. Mech.}{57}{843-852}{2019}



\end{thebibliography}
\end{document}